\documentclass[aps,twocolumn,pra,superscriptaddress,showpacs,tightenlines]{revtex4}
\usepackage{amssymb}
\usepackage{amsmath}
\usepackage{graphicx}
\usepackage{epsfig}
\usepackage{txfonts}
\usepackage{subfigure}
\usepackage{amsfonts}

\begin{document}
\title{Controlling the transport of single photons by tuning the frequency of
either one or two cavities in an array of coupled cavities}
\author{Jie-Qiao Liao}
\affiliation{Advanced Science Institute, The Institute of Physical
and Chemical Research (RIKEN), Wako-shi 351-0198, Japan}
\affiliation{Institute of Theoretical Physics, Chinese Academy of
Sciences, Beijing 100190, China}
\author{Z. R. Gong}
\affiliation{Advanced Science Institute, The Institute of Physical
and Chemical Research (RIKEN), Wako-shi 351-0198, Japan}
\affiliation{Institute of Theoretical Physics, Chinese Academy of
Sciences, Beijing 100190, China}
\author{Lan Zhou}
\affiliation{Key Laboratory of Low-Dimensional Quantum Structures
and Quantum Control of Ministry of Education, Hunan Normal
University, Changsha 410081, China and Department of Physics, Hunan
Normal University, Changsha 410081, China}
\author{Yu-xi Liu}
\affiliation{Advanced Science Institute, The Institute of Physical
and Chemical Research (RIKEN), Wako-shi 351-0198, Japan}
\affiliation{Institute of Microelectronics, Tsinghua University,
Beijing 100084, China}
\author{C. P. Sun}
\affiliation{Advanced Science Institute, The Institute of Physical
and Chemical Research (RIKEN), Wako-shi 351-0198, Japan}
\affiliation{Institute of Theoretical Physics, Chinese Academy of
Sciences, Beijing 100190, China}
\author{Franco Nori}
\affiliation{Advanced Science Institute, The Institute of Physical
and Chemical Research (RIKEN), Wako-shi 351-0198, Japan}
\affiliation{Physics Department, The University of Michigan, Ann
Arbor, MI 48109-1040, USA.}

\date{\today}

\begin{abstract}
We theoretically study how to control transport, bound states, and
resonant states of a single photon in a one-dimensional
coupled-cavity array. We find that the transport of a single photon
in the cavity array can be controlled by tuning the frequency of
either one or two cavities. If one of the cavities in the array has
a tunable frequency, and its frequency is tuned to be larger (or
smaller) than those of other cavities, then there is a photon bound
state above (or below) the energy band of the coupled-cavity array.
However, if two cavities in the array have tunable frequencies, then
there exist both bound states and resonant states. If the
frequencies of the two cavities are chosen to be much larger than
those of other cavities, and the hopping couplings between any two
nearest-neighbor cavities are weak, then a single photon with a
resonant wave vector can be trapped in the region between the two
frequency-tunable cavities. In this case, a quantum supercavity can
be formed by these two frequency-tunable cavities. We also study how
to apply this photon transport control to an array of coupled
superconducting transmission line resonators.
\end{abstract}
\pacs{03.67.Hk, 03.65.-w, 05.60.Gg}

\maketitle \narrowtext
\section{\label{Sec:1}Introduction}
In a quantum network based on photons~\cite{Kimble}, the nodes can
be regarded as information processing stations (e.g., quantum
computers), while the links between any two nodes are provided by
the information carriers (e.g., photons). Due to the high-speed
transmission and low dissipation in optical fibers, photons are
considered to be excellent information carriers (both for classical
and quantum transmissions).

In recent years, with the development of nano-optics, numerous
photonic information processing proposals have been implemented by
using on-chip solid state devices, such as semiconducting
microcavities (e.g., Ref.~\cite{Vahala}) and superconducting
transmission line resonators (e.g.,
Refs.~\cite{youreview,tlrreview,Wendinreview,Wallraff,Rakhmanov,Tsomokos}).
Therefore, how to realize on-chip single-photon devices (e.g.,
Refs.~\cite{Lukin-nature,Liu,you07}) becomes now an increasingly
important research area. For example, single-photon switches (e.g.,
Refs.~\cite{Sun,Switch,ZGLSN08,GIZS08,Liao09}), which control
single-photon transport at will (e.g.,
Refs.~\cite{Lukin-np,Fanpapers}), play an important role in this
area.

Coupled-cavity arrays
(CCAs)~\cite{HBP06,GTCH06,RF07,ASB07,BAB07,ASYE07,HLSS08,ZLS07,ZDLSN08,HZSS07,ZGSS08,tshi}
are one type of photonic system, which has been proposed to process
photonic quantum information. Compared with the usual waveguides,
which have only a linear dispersion relation, the \textit{nonlinear
dispersion} of CCAs can result in the emergence of bound states of
single photons. Many proposals have been put forward to realize
quantum switches in CCAs, which could be used to control
single-photon transport. For example, the reflection and
transmission of single photons in a one-dimensional coupled
resonator waveguide can be controlled by a tunable two-level system
inside one of the cavities~\cite{ZGLSN08}. Moreover, controllable
single-photon transport in a one-dimensional CCA with a tunable
hopping coupling has recently been studied~\cite{Liao09}.

In this paper, we study another approach to realize controllable
single-photon transport in a one-dimensional CCA, considering either
one \textit{frequency-tunable cavity} (FTC) or two FTCs. This work
is motivated by recent mostly experimental results on
frequency-tunable transmission line resonators (e.g.,
Refs.~\cite{Wendin,Tsai,Lehnertnp,Lehnert,Esteve,Delsing,Johansson}),
where the frequencies of the resonators can be changed by varying
either the boundary condition of the electromagnetic wave or the
magnetic flux through the SQUIDs used to construct the transmission
line resonators. In contrast to Refs.~\cite{ZGLSN08,ZDLSN08}, here
the photon transport is controlled by \textit{tuning the frequency
of the cavity}, and there is \textit{no} additional two-level
system, placed inside one of the cavities, to control photon
transport. Therefore, this proposal seems to be simpler and easier
to implement experimentally than those in
Refs.~\cite{ZGLSN08,ZDLSN08}. By changing the frequency of either
one FTC or two FTCs, the reflection and transmission of a single
photon in the coupled-cavity array can be controlled. We also study
the photon bound states and photon resonant states~\cite{scat-pole}
in this coupled-cavity array.

For the coupled-cavity array with \textit{one} frequency-tunable
cavity, if the frequency of the frequency-tunable cavity is larger
than that of other cavities, there is a \textit{bound} state above
the energy band of the corresponding bosonic tight-binding
model~\cite{Data}; while the bound state is below the energy band
when the frequency of the frequency-tunable cavity is smaller than
that of other cavities.

For the CCA with \textit{two} FTCs, we find that there exist
\textit{bound} states around the FTCs. Moreover, when the
frequencies of the two FTCs are much larger than those of other
cavities and the hopping couplings between any two nearest-neighbor
cavities are weak, for resonant wave vectors, a single photon can be
in resonance with the CCA and then remain trapped in the cavities
between the two FTCs. A single photon in \textit{resonance} with the
CCA behaves as a photon inside a supercavity~\cite{ZDLSN08}.

This paper is organized as follows: In Sec.~\ref{Sec:2}, we study
controllable single-photon transport and bound states in a CCA with
one FTC. In Sec.~\ref{Sec:3}, we study controllable single-photon
transport, bound states, and resonant states in the CCA with two
FTCs. In Sec.~\ref{Sec:4}, we present a possible experimental
implementation of our proposal using superconducting transmission
line resonators. A summary is given in Sec.~\ref{Sec:5}.

\section{\label{Sec:2}Coupled cavity array with one frequency-tunable cavity}
\begin{figure}[tbp]
\includegraphics[bb=64 178 524 644, width=8 cm]{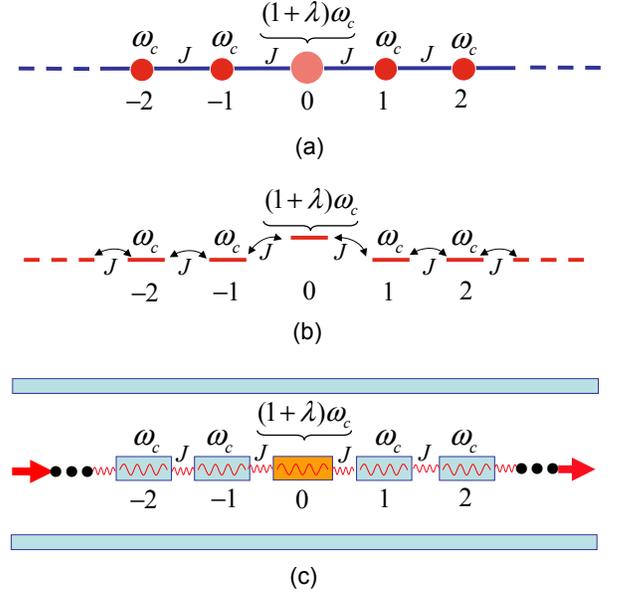}
\caption{(Color online) Schematic diagram of a one-dimensional
coupled-cavity array (CCA): (a) a lattice model for the CCA. Each
disk represents a cavity (labeled by the integer below each dot)
with frequency shown right above each cavity. The larger disk
(located at the site $j=0$) represents the frequency-tunable cavity;
(b) the energy levels of the CCA; (c) coupled superconducting
transmission line resonator array. The central cavity located at the
site $j=0$ is shown in orange. This central cavity has a tunable
frequency $(1+\lambda)\omega_{c}$, where $\lambda$ is the detuning
parameter used to control the transport of single photons through
this coupled-cavity array. If $\lambda>0$ ($\lambda<0$), then there
is a photon bound state above (below) the energy band of the
coupled-cavity array. Thus, the central cavity acts as a tunable
``impurity". The arrows on the far left (right) schematically
indicate the incoming (outgoing) photons. The blue dots represent
the remaining cavities which are not shown here. The integers below
each cavity label each one of them. }\label{configuration1}
\end{figure}

As schematically shown in Fig.~\ref{configuration1}, we consider a
one-dimensional coupled-cavity array, which consists of a chain of
$N$ cavities. We assume that $N$ is a large enough number, so
periodic boundary conditions become reasonable. For specificity, and
without loss of generality, we assume that $N$ is an odd number. The
distance between any two nearest-neighbor cavities is $d_{0}$. The
cavities, except the central one (i.e., the $0$th cavity), have the
same resonant frequency $\omega_{c}$. The central $0$th cavity has
the resonant frequency $(1+\lambda)\omega_{c}$, where $\lambda$ is
used to characterize the detuning between the $0$th cavity and other
identical cavities, and assume to be varied for controlling the
photon transport properties in this system. The frequency of the
$0$th cavity can be larger ($\lambda>0$) or smaller ($\lambda<0$)
than those of other cavities. Any two nearest-neighbor cavities are
coupled via a homogeneous hopping interaction of strength $J$. The
Hamiltonian (with $\hbar=1$) of the CCA reads
\begin{equation}
\hat{H}^{(1)}_{\textrm{CCA}}=\lambda\omega_{c}\hat{a}_{0}^{\dag}\hat{a}_{0}
+\sum_{j=-\frac{N-1}{2}}^{\frac{N-1}{2}}\omega_{c}\hat{a}_{j}^{\dag}\hat{a}_{j}
-\sum_{j=-\frac{N-1}{2}}^{\frac{N-1}{2}}J(\hat{a}_{j}^{\dag}\hat{a}_{j+1}+\hat{a}_{j+1}^{\dag}\hat{a}_{j}),\label{Hamiltonian}
\end{equation}
where $\hat{a}_{j}$ $(\hat{a}^{\dag}_{j})$ is the annihilation
(creation) operator of the $j$th cavity. The superscript ``1" in
$\hat{H}^{(1)}_{\textrm{CCA}}$ denotes that the CCA contains one
frequency-tunable cavity. The first two terms in
Eq.~(\ref{Hamiltonian}) are the ``free Hamiltonian" of the CCA,
while the last term in Eq.~(\ref{Hamiltonian}) represents the
hopping interaction, with strength $J$, between any two
nearest-neighbor cavities. For instance, the term
$\hat{a}_{j}^{\dag}\hat{a}_{j+1}$ means that a photon is annihilated
in the $(j+1)$th cavity and another photon is created in the $j$th
cavity. Hereafter, we only use $\sum_{j}$ instead of the sum shown
in Eq.~(\ref{Hamiltonian}). Since the frequency,
$(1+\lambda)\omega_{c}$, of the $0$th cavity should be nonnegative,
then $\lambda\geq-1$. For $\lambda=0$, the above
Hamiltonian~(\ref{Hamiltonian}) reduces to the usual bosonic
tight-binding (btb) Hamiltonian
\begin{equation}
\hat{H}_{\textrm{btb}}=\omega_{c}\sum_{j}\hat{a}_{j}^{\dag}\hat{a}_{j}
-J\sum_{j}(\hat{a}_{j}^{\dag}\hat{a}_{j+1}+\hat{a}_{j+1}^{\dag}\hat{a}_{j}),\label{bosonictightbinding}
\end{equation}
which can be diagonalized
\begin{equation}
\hat{H}_{\textrm{btb}}=\sum_{k}\Omega_{k}\hat{a}_{k}^{\dag}\hat{a}_{k}
\end{equation}
by using the discrete Fourier transform
\begin{equation}
\hat{a}_{k}=\frac{1}{\sqrt{N}}\sum_{j}\exp(ikjd_{0})\,\hat{a}_{j}\label{Fouriertransform}
\end{equation}
and the periodic boundary conditions, where
\begin{equation}
\Omega_{k}=\omega_{c}-2J\cos (kd_{0})
\end{equation}
is a \textit{nonlinear dispersion relation}, which is an energy
band. Hereafter, the distance $d_{0}$ between two neighboring
cavities is scaled as unity, and $k=2\pi n_{k}/N$  (with $-N/2 <
n_{k}\leq N/2$) are the photon wave vectors.

\subsection{Controllable single-photon transport for one frequency-tunable cavity}
Since the total excitation number operator
$\hat{N}\equiv\sum_{j}\hat{a}_{j}^{\dag}\hat{a}_{j}$ of the CCA is a
conserved observable, i.e.,
$[\hat{N},\hat{H}^{(1)}_{\textrm{CCA}}]=0$,  it is reasonable to
restrict our discussions to the single-particle excitation subspace
for studying single-photon transport. A general state in the
single-excitation subspace can be written as
\begin{equation}\label{eq:6}
|\omega\rangle=\sum_{j}c_{j}|1_{j}\rangle,
\end{equation}
where the state
$|1_{j}\rangle=|0\rangle\otimes\,\cdots\,\otimes|1\rangle_{j}
\otimes\,\cdots\,\otimes|0\rangle$
represents the case when the $j$th cavity has one photon, while
other cavities have no photons. Also, $c_{j}$ is the probability
amplitude of the state $|1_{j}\rangle$. Using the discrete
scattering method studied in Ref.~\cite{ZGLSN08} and the
eigenequation
$\hat{H}^{(1)}_{\textrm{CCA}}|\omega\rangle=\omega|\omega\rangle$,
we obtain
\begin{equation}
\label{eq:cjequation}-J(c_{j+1}-c_{j-1})=[\omega-(1+\lambda\delta_{j0})\omega_{c}]c_{j},
\end{equation}
where $\delta_{j0}$ is the Kronecker delta function.

\begin{figure}[tbp]
\includegraphics[bb=96 272 517 634,width=7.5 cm]{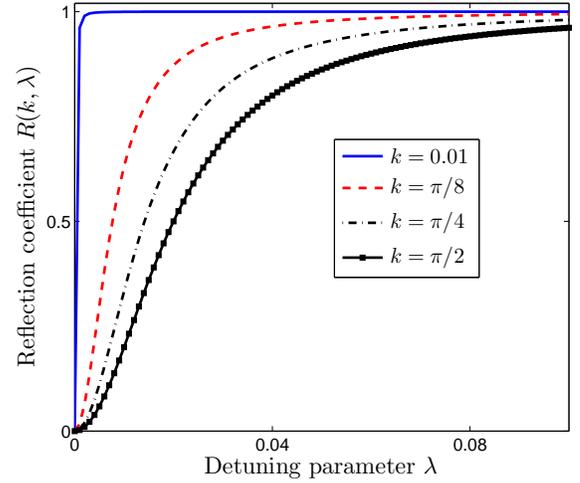}
\caption{(Color online) The photon reflection coefficient
$R(k,\lambda)$ versus the detuning parameter $\lambda$ of the $0$th
cavity is plotted for $k=0.01,\pi/8,\pi/4$ and $\pi/2$, where the
parameters are taken in units of $\omega_{c}$, and
$J/\omega_{c}=0.01$. Recall that the on-site detuning of the central
($0$th) frequency-tunable cavity is $\lambda\omega_{c}$. A
relatively small amount of detuning can make the reflection
coefficient near one.}\label{reflection1}
\end{figure}

Without loss of generality, we assume that a single photon with
frequency $\omega=\Omega_{k}$ is injected from the left side of the
CCA, and then the photon probability amplitudes $c_{j}$ are assumed
to have the following solutions,
\begin{eqnarray}
c_{j}=\left\{
\begin{array}{c}
e^{ikj}+re^{-ikj},\hspace{0.5 cm}j<0, \\
se^{ikj},\hspace{1.5 cm}j>0,
\end{array}
\right.\label{eq:cjrelation}
\end{eqnarray}
where $r$ and $s$ are the photon reflection and transmission
amplitudes, respectively. Here ``$i$" denotes the imaginary unit,
except when specified otherwise. It is easy to check that
Eq.~(\ref{eq:cjrelation}) is the solution of
Eq.~(\ref{eq:cjequation}) when $j\neq0$. Connecting
Eq.~(\ref{eq:cjequation}) at $j=0$ with the continuity condition,
$1+r=s$, for the wave function, we obtain the photon reflection
amplitude
\begin{equation}
r=\frac{\lambda\omega_{c}}{2iJ\sin k-\lambda\omega_{c}},
\end{equation}
which leads to the photon reflection coefficient
\begin{eqnarray}
R(k,\lambda)\equiv|r|^{2}=\frac{(\lambda\omega_{c})^{2}}{4J^{2}\sin^{2}k+(\lambda\omega_{c})^{2}}.\label{reflectioncoefficient}
\end{eqnarray}
The reflection coefficient has three symmetric relations:
$R(k,\lambda)=R(-k,\lambda)$,
$R(\pi/2-k,\lambda)=R(\pi/2+k,\lambda)$, and
$R(k,\lambda)=R(k,-\lambda)$. In Fig.~\ref{reflection1}, the
reflection coefficient $R(k,\lambda)$, as a function of the detuning
parameter $\lambda$, is plotted for $k=0.01,\pi/8,\pi/4$, and
$\pi/2$. It can be seen from Fig.~\ref{reflection1} that the photon
\textit{reflection coefficient} $R(k,\lambda)$ \textit{can be tuned}
from zero to one by changing the detuning parameter $\lambda$.

\subsection{Bound states for one frequency-tunable cavity}
\begin{figure}[tbp]
\includegraphics[bb=89 161 498 689, width=7.8 cm]{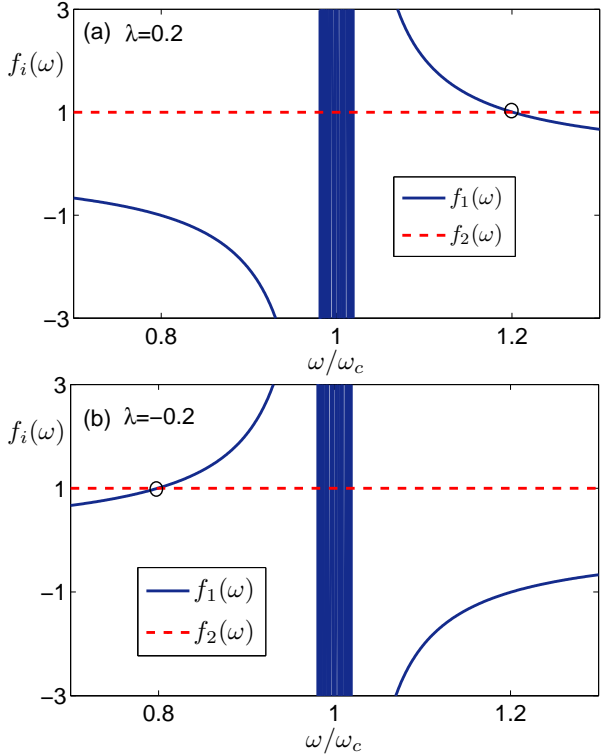}
\caption{(Color online) The functions
$f_{1}(\omega)=\lambda\omega_{c}\sum_{k}[1/(\omega-\Omega_{k})]/N$
(blue) and $f_{2}(\omega)=1$ (red) versus the scaled frequency
$\omega/\omega_c$. The values of $\omega$ corresponding to the
crossing points between the curves for the functions $f_{1}(\omega)$
and $f_{2}(\omega)$ are the frequencies $\omega_{k}$. In other
words, the $\omega_{k}$'s satisfy
$f_{1}(\omega_{k})=f_{2}(\omega_{k})$. Here, the parameters are
chosen as $N=21$, $\omega_{c}=1$, $J/\omega_{c}=0.01$. Note that (a)
and (b) use $\lambda=0.2$ and $\lambda=-0.2$, respectively, and this
is their only difference. Obviously, for $\lambda=0.2$
($\lambda=-0.2$), there is a bound state (shown as a black circle)
above (below) the energy band
($0.98<\omega/\omega_c<1.02$).}\label{localizedmodes}
\end{figure}

A bound state may be formed when a particle is scattered by a
localized potential. In the present model, the on-site extra energy
$\lambda\omega_{c}$ acts as a potential. Therefore, a priori, this
system may have bound states. First, we give a qualitative analysis
of the bound states in the CCA. We now apply the discrete Fourier
transform, defined in Eq.~(\ref{Fouriertransform}), to express the
Hamiltonian~(\ref{Hamiltonian}) in wave vector space as follows,
\begin{equation}
\hat{H}^{(1)}_{\textrm{CCA}}=\sum_{k}\Omega_{k}\hat{a}^{\dag}_{k}\hat{a}_{k}
+\frac{\lambda\omega_{c}}{N}\sum_{k,k'}\hat{a}^{\dag}_{k}\hat{a}_{k'}.\label{Hinkspace}
\end{equation}
In terms of the Hamiltonian in Eq.~(\ref{Hinkspace}), we obtain the
Heisenberg's equation of motion for the operator $\hat{a}_{k}$
\begin{eqnarray}
i\dot{\hat{a}}_{k}=\Omega_{k}\hat{a}_{k}+\frac{\lambda\omega_{c}}{N}\sum_{k'}\hat{a}_{k'}.\label{e12}
\end{eqnarray}
For small $\lambda$, we can assume that
$i\dot{\hat{a}}_{k}=\omega_{k}\hat{a}_{k}$. By introducing the
operator $\hat{b}\equiv\sum_{k}\hat{a}_{k}$, we obtain
\begin{eqnarray}
\hat{a}_{k}=\frac{\lambda\omega_{c}}{N}\frac{1}{\omega_{k}-\Omega_{k}}\hat{b}.\label{e13}
\end{eqnarray}
Making the summation
\begin{eqnarray}
\hat{b}=\sum_{k}\hat{a}_{k}=\frac{\lambda\omega_{c}}{N}\sum_{k}\frac{1}{\omega_{k}-\Omega_{k}}\hat{b},\label{e14}
\end{eqnarray}
then the frequencies $\omega_{k}$ are determined by the equation
\begin{eqnarray}
\frac{\lambda\omega_{c}}{N}\sum_{k}\frac{1}{\omega_{k}-\Omega_{k}}=1.\label{eq:e10}
\end{eqnarray}
Equation~(\ref{eq:e10}) can be solved numerically. In
Figs.~\ref{localizedmodes}, the functions
$f_{1}(\omega)=\lambda\omega_{c}\sum_{k}[1/(\omega-\Omega_{k})]/N$
and $f_{2}(\omega)=1$ are plotted for $\lambda>0$ and $\lambda<0$.
The values of $\omega$ corresponding to the crossing points of both
curves $f_{1}(\omega)$ and $f_{2}(\omega)$ are the solutions
$\omega_{k}$ that satisfy Eq.~(\ref{eq:e10}). Obviously, when
$\lambda>0$ there is a bound state above the energy band, while for
$\lambda<0$ there is a bound state below the energy band. These
bound states are shown as black circles in Fig.~\ref{localizedmodes}

Below, we analytically study the bound states. Since we choose the
frequency-tunable cavity as the coordinate origin, the
Hamiltonian~(\ref{Hamiltonian}) of the system is symmetric around
the $0$th cavity. The eigenstates of the
Hamiltonian~(\ref{Hamiltonian}) have either symmetric or asymmetric
parities. For the asymmetric case, we have the relation
$c_{j}=-c_{-j}$, which implies $c_{0}=0$. Therefore, the frequency
change of the frequency-tunable cavity will not affect the
asymmetric eigenstates. For the symmetry case, we assume the
following solution
\begin{equation}
c_{j}=A\,|\mu|^{j},\label{Bounsoultionform}
\end{equation}
where $\mu$ is a parameter introduced to describe the bound state of
the Hamiltonian~(\ref{Hamiltonian}).

\begin{figure}[tbp]
\includegraphics[bb=65 261 524 611, width=7.2 cm]{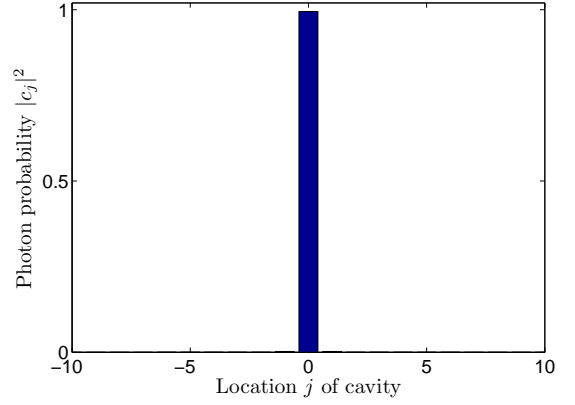}
\caption{(Color online) Photon probability $|c_{j}|^{2}$ versus the
location $j$ of the cavity. The $c_{j}$'s are introduced in
Eq.~(\ref{eq:6}). Other parameters are taken as $\omega_{c}=1$,
$|\lambda|=0.2$, and $J/\omega_{c}=0.01$. In this case, since
$|c_{j}|^{2}$ is peaked at the cavity located at $j=0$, then
\textit{a single photon is localized} around the central ($0$th)
frequency-tunable cavity.}\label{localizedwave1}
\end{figure}

Substituting the solution~(\ref{Bounsoultionform}) into
Eq.~(\ref{eq:cjequation}), we obtain
\begin{equation}
-J\mu^{2}+\lambda\omega_{c}\mu+J=0.\label{Boundstateeq}
\end{equation}
Equation~(\ref{Boundstateeq}) has two solutions
\begin{equation}
\mu_{\pm}=\frac{-\lambda\omega_{c}\pm\sqrt{4J^{2}+(\lambda\omega_{c})^{2}}}{-2J}.
\end{equation}
When $\lambda>0$, we choose the solution $\mu_{+}$; while for the
case $\lambda<0$ we choose the solution $\mu_{-}$. The corresponding
eigenfrequencies are
\begin{equation}
\omega_{\pm}=\omega_{c}\pm\sqrt{4J^{2}+(\lambda\omega_{c})^{2}}.
\end{equation}
The relations $\omega_{+}>\omega_{c}+2J$ and
$\omega_{-}<\omega_{c}-2J$ mean that the bound state is above and
below the energy band, respectively, as shown in
Figs.~\ref{localizedmodes}(a) and~~\ref{localizedmodes}(b). These
analytical results are consistent with our previous analysis in
Eqs.~(\ref{Hinkspace}--\ref{eq:e10}) and Fig.~\ref{localizedmodes}.
Note that $\mu_{+}(|\lambda|)=-\mu_{-}(-|\lambda|)$, so
$|c_{j}|^{2}$ is the same for the two bound states. In
Fig.~\ref{localizedwave1}, we plot $|c_{j}|^{2}$ as a function of
the lattice parameter $j$. Figure~\ref{localizedwave1} shows that a
single photon is mainly localized around the central ($0$th)
frequency-tunable cavity.

\subsection{Links to localized excitations in solids}

Periodic solid state systems exhibit bands.  Adding localized
defects to a translationally invariant structure induces localized
states around those defects. In general, adding a defect (i.e.,
anything that breaks translational symmetry, like an impurity, or an
interface) is enough to create gap states outside the continuous
bands.  Thus, gap states outside the bands are linked to localized
states.

Figure~\ref{localizedmodes} in this work is related to Fig.~9.9 in
page 395 of Madelung's classic textbook~\cite{Madelung} on solid
state physics. Indeed, equation~(9.37) in~\cite{Madelung} is related
to Eq.~(\ref{eq:e10}) in our paper. The links between them is that a
defect added to a periodic structure tends to localize excitations
around the defect, and this localized state corresponds to gap
states.

\section{\label{Sec:3}Coupled cavity array with two frequency-tunable cavities}
\begin{figure}[tbp]
\includegraphics[bb=50 190 540 672, width=7.5 cm]{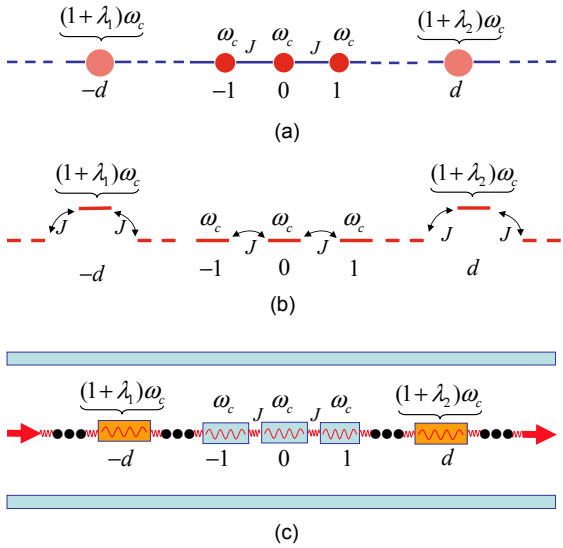}
\caption{(Color online) Schematic configuration of a one-dimensional
coupled-cavity array with two frequency-tunable cavities: (a) a
lattice model for the CCA. Each disk represents a cavity (labeled by
the integer below each cavity) and with frequency shown right above
it. The two larger disks (located at the sites $j=\pm d$) represent
the two frequency-tunable cavities; (b) the energy levels of the
CCA; (c) schematic diagram of a coupled superconducting transmission
line resonator array. The two detuned cavities have frequencies
$(1+\lambda_{1})\omega_{c}$ and $(1+\lambda_{2})\omega_{c}$. The two
detuning parameters, $\lambda_{1}$ and $\lambda_{2}$, control the
transport of single photons in the array of coupled cavities. The
incoming photon can be localized at the cavities located at $j=\pm
d$, forming localized bound states (in
Fig.~\ref{boundstateprobabilityfortwoftc}). Alternatively, when the
photon wave vector is in resonance, the photon can be confined in
the region between these two cavities, as shown in
Fig.~\ref{resonantstateprobability}}\label{configuration2}
\end{figure}

As schematically shown in Fig.~\ref{configuration2}, we now consider
the case when there are \textit{two} FTCs in the CCA.  These two
FTCs are located in the $-d$th and the $d$th cavities, respectively.
The Hamiltonian can now be written as
\begin{eqnarray}
\hat{H}^{(2)}_{\textrm{CCA}}&=&\lambda_{1}\omega_{c}\hat{a}_{-d}^{\dag}\hat{a}_{-d}+\lambda_{2}\omega_{c}\hat{a}_{d}^{\dag}\hat{a}_{d}
+\omega_{c}\sum_{j}\hat{a}_{j}^{\dag}\hat{a}_{j}\nonumber\\
&&-J\sum_{j}(\hat{a}_{j}^{\dag}\hat{a}_{j+1}+\hat{a}_{j+1}^{\dag}\hat{a}_{j}),\label{oHamiltonian}
\end{eqnarray}
where $\lambda_{1}$ and $\lambda_{2}$ are, respectively, used to
describe the frequency of the $-d$th and the $d$th cavities, and $J$
is the hopping coupling strength between any two nearest-neighbor
cavities. The superscript ``2" in $\hat{H}^{(2)}_{\textrm{CCA}}$
means that the CCA contains two frequency-tunable cavities.

\subsection{Controllable single-photon transport for two frequency-tunable cavities}
For the CCA with \textit{two} frequency-tunable cavities, the total
excitation number operator
$\hat{N}\equiv\sum_{j}\hat{a}^{\dag}_{j}\hat{a}_{j}$ is also a
conserved observable. Similar to Eq.~(\ref{eq:6}), the eigenstates
of the system can now be written as
$|\omega\rangle=\sum_{j}c_{j}|1_{j}\rangle$, where the coefficients
$c_{j}$ are determined by the equation
\begin{eqnarray}
-J(c_{j-1}+c_{j+1})=\left[\omega-\omega_{c}\left(1+\lambda_{1}\delta_{-d,j}+\lambda_{2}\delta_{d,j}\right)\right]c_{j}.\label{cjequation}
\end{eqnarray}
For simplicity, and without loss of generality, below we assume that
the parameters for the two frequency-tunable cavities are identical,
i.e., $\lambda_{1}=\lambda_{2}=\lambda_{0}$.

\begin{figure}[tbp]
\includegraphics[bb=117 287 503 587, width=7.2 cm]{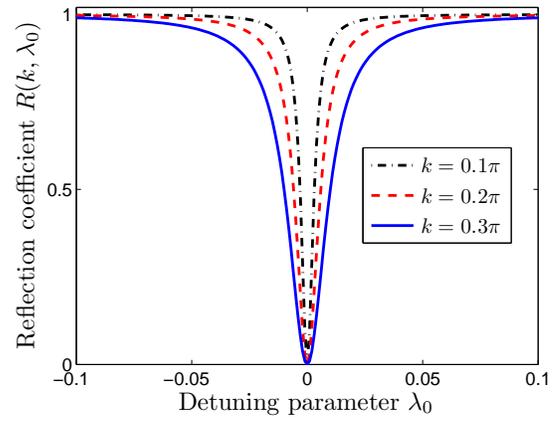}
\caption{(Color online) Photon reflection coefficient $R$ plotted
versus the detuning parameter $\lambda_{0}$ for $k=0.1\pi$,
$0.2\pi$, and $0.3\pi$, where $\omega_{c}=1$, $d=5$, and
$J/\omega_{c}=0.01$. Here we consider two frequency-tunable cavities
located at $j=-d$ and $j=d$. A relatively small amount of detuning
can make the photon reflection coefficient near
one.}\label{reflection2}
\end{figure}

For a photon with frequency $\omega=\Omega_{k}$, the functions
$\exp(ikj)$ and $\exp(-ikj)$ are the solutions of the
Eq.~(\ref{cjequation}) when $j\neq -d$ and $j\neq d$. Therefore, the
general form of the solution for Eq.~(\ref{cjequation}) is assumed
as
\begin{eqnarray}
c_{j}=\left\{
\begin{array}{c}
e^{ikj}+re^{-ikj},\hspace{1 cm}j<-d, \\
Ae^{ikj}+Be^{-ikj},\hspace{0.5 cm}-d<j<d,\\
se^{ikj},\hspace{2 cm}j>d.
\end{array}
\right.\label{eq:cjform}
\end{eqnarray}
Substituting the solution~(\ref{eq:cjform}) into
Eq.~(\ref{cjequation}) and using the continuity condition at $j=-d$
and $j=d$
\begin{subequations}
\begin{align}
e^{-ikd}+re^{ikd}&=Ae^{-ikd}+Be^{ikd},\\
Ae^{ikd}+Be^{-ikd}&=se^{ikd},\label{continuousequation}
\end{align}
\end{subequations}
we obtain the photon reflection amplitude
\begin{eqnarray}
r&=&\lambda_{0}\omega_{c}
e^{-i(2d-1)k}\left[J\left(1+e^{4ikd}\right)\left(e^{2ik}-1\right)\right.\nonumber\\
&&\left.+\lambda_{0}\omega_{c}\left(e^{4ikd}-1\right)e^{ik}\right]\nonumber\\
&&\times\left\{J\left(e^{2ik}-1\right)\left[J\left(e^{2ik}-1\right)-2\lambda_{0}\omega_{c}e^{ik}\right]\right.\nonumber\\
&&\left.-\lambda_{0}^{2}\omega_{c}^{2}\left(e^{4ikd}-1\right)e^{2ik}\right\}^{-1}.
\end{eqnarray}
In Fig.~\ref{reflection2}, the photon reflection coefficient
$R(k,\lambda_{0})=|r|^2$ is plotted as a function of the parameter
$\lambda_{0}$ for different wave vectors $k=0.1\pi$, $0.2\pi$,
$0.3\pi$. Figure~\ref{reflection2} shows that the reflection
coefficient $R(k,\lambda_{0})$ can be tuned from zero to one by
changing the detuning parameter $\lambda_{0}$.

\subsection{Bound states and resonant states for two frequency-tunable cavities}
To study the bound states of the CCA with two frequency-tunable
cavities, we assume
\begin{eqnarray}
c_{j}=\left\{
\begin{array}{c}
Ae^{-ikj},\hspace{2 cm}j<-d, \\
Be^{ikj}+Ce^{-ikj},\hspace{0.5 cm}-d<j<d,\\
De^{ikj},\hspace{2 cm}j>d.
\end{array}
\right.\label{eq:boundstatecjform}
\end{eqnarray}
From Eq.~(\ref{cjequation}) at $j=-d$, and the condition for the
continuity of the wave function at $j=-d$, (namely:
$Ae^{ikd}=Be^{-ikd}+Ce^{ikd}$), we obtain
\begin{subequations}
\label{avsb2}
\begin{align}
\label{avsb}\frac{A}{B}&=\frac{i2J\sin k}{\lambda_{0}\omega_{c}}\exp(-2ikd),\\
\label{cvsbatmd}\frac{C}{B}&=\frac{(i2J\sin
k-\lambda_{0}\omega_{c})}{\lambda_{0}\omega_{c}}\exp(-2ikd).
\end{align}
\end{subequations}
From Eq.~(\ref{cjequation}) at $j=d$, and the continuity condition
at $j=d$, (that is $Be^{ikd}+Ce^{-ikd}=De^{ikd}$), we obtain
\begin{subequations}
\label{cvsbatd2}
\begin{align}
\label{cvsbatd}\frac{C}{B}&=\frac{\lambda_{0}\omega_{c}}{i2J\sin k-\lambda_{0}\omega_{c}}\exp(2ikd),\\
\label{dvsb}\frac{D}{B}&=\frac{-i2J\sin k}{i2J\sin
k-\lambda_{0}\omega_{c}}.
\end{align}
\end{subequations}
Obviously, the right hand sides of Eqs.~(\ref{cvsbatmd})
and~(\ref{cvsbatd}) should be equal; then we obtain
\begin{eqnarray}
(\lambda_{0}\omega_{c})^{2}
\exp(4ikd)=(\lambda_{0}\omega_{c}-i2J\sin
k)^{2},\label{boundcondition}
\end{eqnarray}
which implies
\begin{eqnarray}
\exp(2ikd)=\pm\left(1-\frac{i2J}{\lambda_{0}\omega_{c}}\sin
k\right).\label{boundconditionpm}
\end{eqnarray}
This Eq.~(\ref{boundconditionpm}) is important here because it
determines the wave vector of the photon states (either bound states
or resonant states). Below, we will discuss the existence of bound
states in this system for the two cases shown in
Eq.~(\ref{boundconditionpm}). These will be denoted as positive root
and negative root, respectively, of Eq.~(\ref{boundcondition}). Also
note that $\lambda_{0}$ takes two values ($\pm|\lambda_{0}|$)
because the frequency detuning of the cavities at $j=\pm d$ can be
either positive or negative.

\begin{figure}[tbp]
\includegraphics[bb=76 206 530 620, width=7.4 cm]{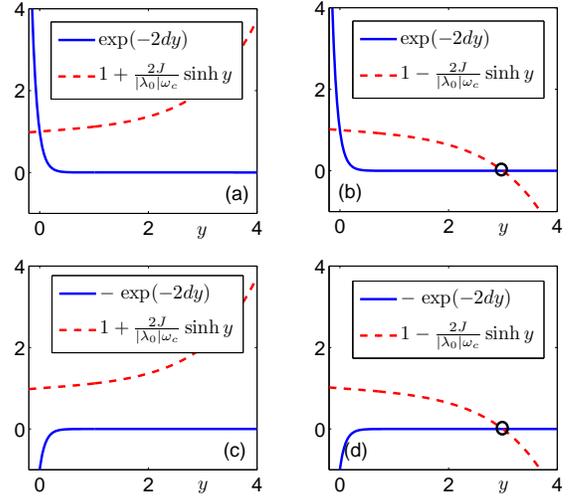}
\caption{(Color online) The functions $\exp(\pm 2dy)$ and
$1\pm2J\sinh y/(|\lambda_{0}|\omega_{c})$ are plotted as a function
of $y$. The wave vector $k=x+iy$, with $y\geq0$ here. The
intersection points shown inside the black circles give the nonzero
solutions of the transcendental equation~(\ref{boundconditionpm}).
The parameters used here are $\omega_{c}=1$, $J/\omega_{c}=0.01$,
$d=5$, and $|\lambda_{0}|=0.2$. Equation~(\ref{boundconditionpm})
determines the wave vector of the photon states (either bound states
or resonant states).}\label{boundstate}
\end{figure}

\subsubsection{Positive root of Eq.~(\ref{boundconditionpm})}
We now assume that the wave vector $k=x+iy$, with $y\geq0$. It
should be pointed out that the parameter $y$ can take the value
$y=0$ to include the possibility for the existence of some
interesting resonant states. Since the energy of the photon must be
a real parameter, (i.e., $\cos k=\cos x\cosh y-i\sin x\sinh y$ is
real), this condition can be satisfied in the following three cases:

(1) $x=2n\pi$, with $n\in Z$ (hereafter, $Z$ denotes the set of
integers). In this case, $y$ is determined by the equation
\begin{eqnarray}
\exp(-2dy)=1+\frac{2J}{\lambda_{0}\omega_{c}}\sinh y.
\label{eqsec5a1}
\end{eqnarray}
The solutions of the transcendent equation~(\ref{eqsec5a1}) (and
other transcendent equations below) are determined through the
numerical method briefly sketched in Fig.~\ref{boundstate}.

\begin{figure}[tbp]
\includegraphics[bb=67 272 494 602, width=6.8 cm]{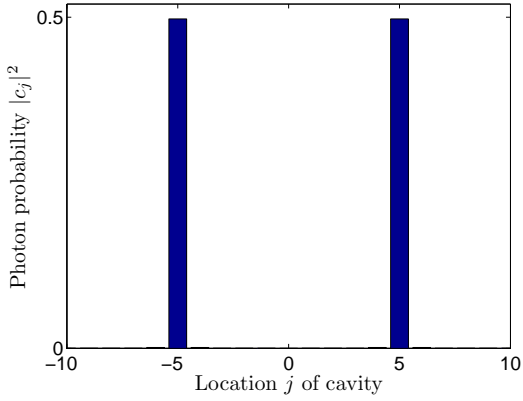}
\caption{(Color online) Photon probability $|c_{j}|^{2}$ of each
coefficient of the normalized wave function given by
Eqs.~(\ref{eq:6}) and ~(\ref{boundsolutionylz26}). The parameters
chosen here are $\omega_{c}=1$, $J/\omega_{c}=0.01$, $d=5$,
$y_{0}$=2.998, and with a detuning $|\lambda_{0}|=0.2$ for the
cavity at $j=\pm d$. The photon is now localized and \textit{bound
at} the two detuned cavities located at $j=\pm
d=\pm5$.}\label{boundstateprobabilityfortwoftc}
\end{figure}

The coefficient relations in Eqs.~(\ref{avsb2}) and (\ref{cvsbatd2})
are
\begin{eqnarray}
\frac{A}{B}=-\frac{D}{B}=\exp(2dy)-1,\hspace{0.5
cm}\frac{C}{B}=-1.\label{coeffrelation1}
\end{eqnarray}
When $\lambda_{0}>0$, from Fig.~\ref{boundstate}(a), we can see that
Eq.~(\ref{eqsec5a1}) has only a ``zero solution", $y=0$, then
$A=D=0$ and $C=-B$, so the corresponding wave function $c_{j}=0$.
When $\lambda_{0}<0$, from Fig.~\ref{boundstate}(b), we know that
Eq.~(\ref{eqsec5a1}) has two solutions: one is zero, and the other
one is a positive number denoted by $y_{0}>0$. For the zero
solution, the wave function is $c_{j}=0$. For the solution
$y_{0}>0$, we have $A=-D=(e^{2dy_{0}}-1)B$ and $C=-B$. The
corresponding wave function~(\ref{eq:boundstatecjform}) becomes
\begin{eqnarray}
c_{j}=\left\{
\begin{array}{c}
B(e^{2dy_{0}}-1)e^{y_{0}j},\hspace{1 cm}j<-d, \\
-2B\sinh(y_{0}j),\hspace{1 cm}-d<j<d,\\
-B(e^{2dy_{0}}-1)e^{-y_{0}j},\hspace{0.7 cm}j>d,
\end{array}
\right.\label{boundsolutionylz26}
\end{eqnarray}
where $B$ is determined by the normalization condition. Obviously,
the wave function~(\ref{boundsolutionylz26}) is asymmetric, i.e.,
$c_{j}=-c_{-j}$, so it is an odd-parity state. Using the parameters
$\omega_{c}=1$, $J/\omega_{c}=0.01$, $d=5$, and $|\lambda_{0}|=0.2$,
we obtain $y_{0}=2.998$. The photon probability corresponding to the
wave function~(\ref{boundsolutionylz26}) is plotted in
Fig.~\ref{boundstateprobabilityfortwoftc}, which shows that the
probability to find a single photon~(\ref{boundsolutionylz26}) is
mostly around the two frequency-tunable cavities.

(2) $x=(2n+1)\pi$, with $n\in Z$. In this case, $y$ is determined by
the equation
\begin{eqnarray}
\exp(-2dy)=1-\frac{2J}{\lambda_{0}\omega_{c}}\sinh
y,\label{eqsec5a27}
\end{eqnarray}
and the coefficient relations in Eqs.~(\ref{avsb2}) and
(\ref{cvsbatd2}) are the same as those in
Eq.~(\ref{coeffrelation1}). When $\lambda_{0}<0$, from
Fig.~\ref{boundstate}(a), we can see that Eq.~(\ref{eqsec5a27}) has
only a zero solution, $y=0$, then $A=D=0$ and $C=-B$. The
corresponding wave function is $c_{j}=0$. When $\lambda_{0}>0$, from
Fig.~\ref{boundstate}(b), it can be seen that Eq.~(\ref{eqsec5a27})
has two solutions, one is zero, and the other is a positive number
denoted by $y_{0}>0$. For the zero solution, the wave function is
$c_{j}=0$. For the solution $y_{0}>0$, then $A=-D=(e^{2dy_{0}}-1)B$
and $C=-B$. The corresponding wave function becomes
\begin{eqnarray} c_{j}=\left\{
\begin{array}{c}
B(e^{2dy_{0}}-1)e^{i(2n+1)\pi j}e^{y_{0}j},\hspace{1 cm}j<-d, \\
-2Be^{i(2n+1)\pi j}\sinh(y_{0}j),\hspace{1 cm}-d<j<d,\\
-B(e^{2dy_{0}}-1)e^{i(2n+1)\pi j}e^{-y_{0}j},\hspace{0.8 cm}j>d.
\end{array}
\right.\label{boundsolutionylz29}
\end{eqnarray}
The wave function in Eq.~(\ref{boundsolutionylz29}) is asymmetric
for $j$, so it is an odd-parity state. The square of the module of
the wave function in Eq.~(\ref{boundsolutionylz29}) is the same as
that of the wave function in Eq.~(\ref{boundsolutionylz26}).

\begin{figure}[tbp]
\includegraphics[bb=81 168 491 674, width=6.8 cm]{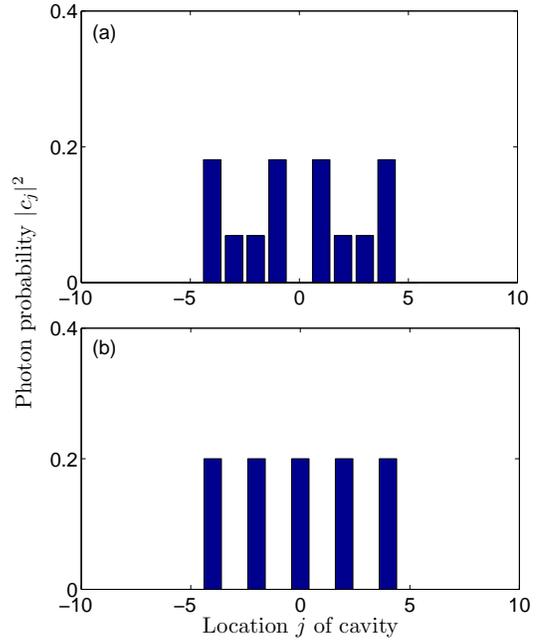}
\caption{(Color online) Photon probability $|c_{j}|^{2}$, given in
Eqs.~(\ref{boundsolutionylz35}) and~(\ref{boundsolutionylz44}),
versus the location $j$ of each cavity are plotted in figures (a)
and (b), respectively. Here the parameters are set as
$\omega_{c}=1$, $m=2$, $d=5$, and $2J/(\lambda_0\omega_c)\approx 0$.
Now the photon is \textit{not bound at} $j=\pm d$, as in
Fig.~\ref{boundstateprobabilityfortwoftc}, but it is
\textit{confined in between} $j=\pm d=\pm5$. This is because a
single photon with wave vector (a) $2\pi/5$ and (b) $\pi/2$ is
\textit{in resonance} with the coupled-cavity array between the two
detuned cavities located at $j=\pm d=\pm5$. These are resonant
states. }\label{resonantstateprobability}
\end{figure}

(3) $y=0$. The motivation for studying the case $y=0$ is to
investigate whether there exist some interesting resonant states.
In this case, $x$ is determined by the equation
\begin{subequations}
\begin{align}
\label{cos}\cos(2dx)&=1,\\
\label{sin}-\frac{2J}{\lambda_{0}\omega_{c}}\sin x&=0.
\end{align}
\end{subequations}
The solution of Eq.~(\ref{cos}) is $x=m\pi/d$, with $m\in Z$. For
Eq.~(\ref{sin}), the solutions are $x=l\pi$, with $l\in Z$, for a
general $2J/(\lambda_{0}\omega_{c})$, or any $x$ when
$\lambda_{0}\omega_{c}\gg2J$, i.e.,
$2J/(\lambda_{0}\omega_{c})\approx0$. Connecting the solutions for
the two Eqs.~(\ref{cos}) and (\ref{sin}), we obtain two solutions
for the case of $y=0$: (a) $x=l\pi$, then the coefficient relations
in Eqs.~(\ref{avsb2}) and (\ref{cvsbatd2}) become
\begin{eqnarray}
\frac{A}{B}=\frac{D}{B}=0,\hspace{0.5
cm}\frac{C}{B}=-1.\label{coeffrelation2}
\end{eqnarray}
Therefore, the wave function is $c_{j}=0$; (b) when
$\lambda_{0}\omega\gg2J$, in this case, we choose $x=m\pi/d$, then
the coefficient relations in Eqs.~(\ref{avsb2}) and (\ref{cvsbatd2})
are the same as those in Eq.~(\ref{coeffrelation2}). The wave
function becomes
\begin{eqnarray} c_{j}=\left\{
\begin{array}{c}
0,\hspace{2 cm}j<-d, \\
2iB\sin\left(\frac{m\pi j}{d}\right),\hspace{1 cm}-d<j<d,\\
0,\hspace{2.2 cm}j>d,
\end{array}
\right.\label{boundsolutionylz35}
\end{eqnarray}
Obviously, the resonant state~(\ref{boundsolutionylz35}) is an
odd-parity state with parameter $j$. The square of the module of the
wave function in Eq.~(\ref{boundsolutionylz35}) is plotted in
Fig.~\ref{resonantstateprobability}(a). We find that, when
$\lambda_{0}\omega_{c}\gg 2J$, (i.e., the frequencies of the $-d$th
and $d$th cavities are very largely detuned from those of other
cavities and the hopping coupling $J$ between two nearest-neighbor
cavities is weak), the photon with wave vector $m\pi/d$  ($m\in Z$)
can produce a \textit{resonance} in the region between the two FTCs,
once it is injected there. Therefore, this resonant photon state is
confined between the two cavities located at $j=\pm d$, as shown in
Fig.~\ref{resonantstateprobability}(a).

\subsubsection{Negative root of Eq.~(\ref{boundconditionpm})}
We now write the wave vector as $k=x+iy$, with $y\geq0$. Since the
energy of the photon must be real, (i.e., $\cos k=\cos x\cosh
y-i\sin x\sinh y$ should be real), this condition can be satisfied
in the following three cases:

(1) $x=2n\pi$. In this case, $y$ is determined by the equation
\begin{eqnarray}
-\,\exp(-2dy)=1+\frac{2J}{\lambda_{0}\omega_{c}}\sinh
y,\label{eqsec5a33}
\end{eqnarray}
and the coefficient relations in Eqs.~(\ref{avsb2}) and
(\ref{cvsbatd2}) become
\begin{eqnarray}
\frac{A}{B}=\frac{D}{B}=e^{2dy}+1,\hspace{0.5
cm}\frac{C}{B}=1.\label{coeffrelation3}
\end{eqnarray}
In this case, when $\lambda_{0}>0$, we know from
Fig.~\ref{boundstate}(c) that Eq.~(\ref{eqsec5a33}) has no positive
solution. When $\lambda_{0}<0$, from Fig.~\ref{boundstate}(d), we
know that Eq.~(\ref{eqsec5a33}) has a positive solution $y_{1}>0$.
Then $A=D=(e^{2dy_{1}}+1)B$ and $C=B$. Therefore, the wave function
becomes
\begin{eqnarray} c_{j}=\left\{
\begin{array}{c}
B(e^{2dy_{1}}+1)e^{y_{1}j},\hspace{1 cm}j<-d, \\
2B\cosh(y_{1}j),\hspace{1 cm}-d<j<d,\\
B(e^{2dy_{1}}+1)e^{-y_{1}j},\hspace{1 cm}j>d.
\end{array}
\right.\label{boundsolutionylz38}
\end{eqnarray}
The wave function in Eq.~(\ref{boundsolutionylz38}) is an
even-parity state. Using the parameters $\omega_{c}=1$,
$J/\omega_{c}=0.01$, $d=5$, and $|\lambda_{0}|=0.2$, we obtain
$y_{1}=2.998$. We found that the difference between $y_{0}$ and
$y_{1}$ is very small, on the order of $10^{-12}$. Thus the photon
probability $|c_{j}|^{2}$ here looks like the one in
Fig.~\ref{boundstateprobabilityfortwoftc}.

(2) $x=(2n+1)\pi$. In this case, $y$ is determined by the equation
\begin{eqnarray}
-\,\exp(-2dy)=1-\frac{2J}{\lambda_{0}\omega_{c}}\sinh
y,\label{eqsec5a36}
\end{eqnarray}
and the coefficient relations in Eqs.~(\ref{avsb2}) and
(\ref{cvsbatd2}) are the same as those in
Eq.~(\ref{coeffrelation3}). When $\lambda_{0}<0$, from
Fig.~\ref{boundstate}(c), we can find that the Eq.~(\ref{eqsec5a36})
has no positive solution. When $\lambda_{0}>0$, from
Fig.~\ref{boundstate}(d), we know that Eq.~(\ref{eqsec5a36}) has one
positive solution $y_{1}$. Then $A=D=(e^{2dy_{1}}+1)B$ and $C=B$. So
the corresponding wave function now becomes
\begin{eqnarray} c_{j}=\left\{
\begin{array}{c}
B(e^{2dy_{1}}+1)e^{i(2n+1)\pi j}e^{y_{1}j},\hspace{1 cm}j<-d, \\
2Be^{i(2n+1)\pi j}\cosh(y_{1}j),\hspace{1 cm}-d<j<d,\\
B(e^{2dy_{1}}+1)e^{i(2n+1)\pi j}e^{-y_{1}j},\hspace{1 cm}j>d,
\end{array}
\right.\label{boundsolutionylz47}
\end{eqnarray}
This wave function~(\ref{boundsolutionylz47}) is an even-parity
state. The square of the module of the wave function in
Eq.~(\ref{boundsolutionylz47}) is the same as that of the wave
function in Eq.~(\ref{boundsolutionylz38}).

(3) $y=0$. In this case, $x$ is determined by the equation
\begin{subequations}
\begin{align}
\label{cos1}\cos(2dx)&=-1,\\
\label{sin1}\frac{2J}{\lambda_{0}\omega_{c}}\sin x&=0.
\end{align}
\end{subequations}
The solutions of Eq.~(\ref{cos1}) are $x=(2m+1)\pi/(2d)$, and the
solutions of Eq.~(\ref{sin1}) are $x=l\pi$ or any $x$ when
$\lambda_{0}\omega_{c}\gg2J$. Therefore, the solutions meeting the
two equations~(\ref{cos1}) and~(\ref{sin1}) at the same time are
$x=(2m+1)\pi/(2d)$, when $\lambda_{0}\omega_{c}\gg2J$. Then the
coefficient relations in Eqs.~(\ref{avsb2}) and (\ref{cvsbatd2})
become
\begin{eqnarray}
\frac{A}{B}=\frac{D}{B}=0,\hspace{0.5 cm}\frac{C}{B}=1.
\end{eqnarray}
Therefore, the wave function now becomes
\begin{eqnarray} c_{j}=\left\{
\begin{array}{c}
0,\hspace{2 cm}j<-d, \\
2B\cos\left(\frac{(2m+1)\pi j}{2d}\right),\hspace{0.5 cm}-d<j<d,\\
0,\hspace{2 cm}j>d.
\end{array}
\right.\label{boundsolutionylz44}
\end{eqnarray}
This resonant state~(\ref{boundsolutionylz44}) is an even-parity
state. The square of the module of the wave
function~(\ref{boundsolutionylz44}) is plotted in
Fig.~\ref{resonantstateprobability}(b). This figure shows that a
single photon with wave vectors $(2m+1)\pi/(2d)$ can be in resonance
in the region between the two frequency-tunable cavities. For the
two cases of resonant states given in
Eqs.~(\ref{boundsolutionylz35}) and~(\ref{boundsolutionylz44}), the
center cavities between the two frequency-tunable cavities form a
supercavity~\cite{ZDLSN08}. We note that our approach is also valid
for the case $\lambda_{1}\neq\lambda_{2}$.

\section{\label{Sec:4}Physical realization of frequency-tunable superconducting transmission line resonators}
In this section, we study several physical realizations of a
frequency-tunable coupled-cavity array by using superconducting
transmission line resonators. In recent years, there have been
several theoretical proposals and experiments on how to realize a
frequency-tunable transmission line resonator (e.g.,
Refs.~\cite{Wendin,Tsai,Lehnertnp,Lehnert,Esteve,Delsing,Johansson}).

Typically, there are two physical mechanisms to tune the resonant
frequency of a superconducting transmission line resonator. One
method is to change the boundary condition of the electromagnetic
wave in a transmission line. By changing the boundary condition, the
effective wavelengths (also effective frequencies) of the resonant
modes are changed~\cite{Wendin,Delsing,Tsai}.

Another method is to construct a transmission line resonator by
using a series of magnetic-flux-biased SQUIDs. Since the effective
inductor of a magnetic-flux-biased SQUID can be tuned by changing
the biased magnetic flux~\cite{Lehnertnp,Lehnert,Esteve}, the
inductance per unit length of the SQUID array is controllable.
Therefore, the resonant frequencies of the modes in the SQUID array
can be tuned by controlling the biased magnetic flux threading
through the SQUIDs.

Below, we present a brief review of these two methods already used
to obtain frequency-tunable transmission line resonator. The
original derivations of the two methods have been given in
Refs.~\cite{Wendin,Lehnert}, but for the sake of completeness of
this paper, here we briefly review the main aspects of these.

\subsection{Tuning the frequency of a superconducting transmission line resonator: changing the boundary condition}
\begin{figure}[tbp]
\includegraphics[bb=113 430 450 625, width=7.8 cm]{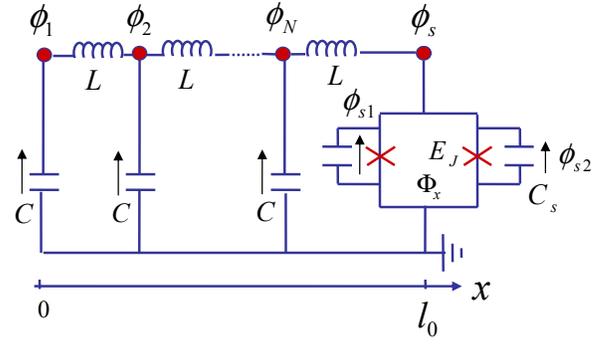}
\caption{(Color online) Circuit model of a frequency-tunable
transmission line resonator integrated with a SQUID~\cite{Wendin}.}
\label{fretunableTLR.eps}
\end{figure}

We briefly summarize the mechanism for frequency tunability of a
superconducting transmission line resonator by controlling its
boundary condition~\cite{Wendin}. The lumped element circuit of a
superconducting transmission line resonator with a symmetric SQUID,
which is equivalent to a chain of identical $LC$ circuits, is shown
in Fig.~\ref{fretunableTLR.eps}.
Here, $\phi_{j}$ is the phase variable of the $j$th node; $C$ and
$L$ are respectively the capacitance and inductance of each $LC$
circuit; $\phi_{s1}$ and $\phi_{s2}$ are the phase variables across
the left and right Josephson junctions in the SQUID, respectively;
$C_{s}$ is the capacitance of one junction in the SQUID. The SQUID
is equivalent to a junction with an effective Josephson energy
$E_{J}(f)=2E_{J}\cos \left(f/2\right)$. Here $E_{J}$ is the
Josephson energy of one junction, $f=2\pi\Phi_{x}/\Phi_{0}$, where
$\Phi_{x}$ is the flux through the loop of the SQUID and $\Phi_{0}$
is the magnetic flux quanta. $\phi _{s}=(\phi _{s1}+\phi _{s2})/2$
is the net phase across the SQUID. Note that here the
self-inductance of the superconducting loop is neglected. When $\phi
_{s}\ll 1$, and the charging energy and Josephson energy satisfy the
condition $E_{J}(f)\gg C_{s}[\Phi_{0}/(2\pi)]^{2}$, then the SQUID
can be approximated as a harmonic oscillator~\cite{Wendin}. For the
phase variable of the transmission line resonator, the wave equation
reads~\cite{Wendin}
\begin{equation}
\ddot{\phi}\left(x,t\right)-v^{2}\phi ^{\prime \prime }\left(
x,t\right) =0,\label{waveequationofTLR}
\end{equation}
where $v=1/\sqrt{C_{0}L_{0}}$. Here $C_{0}$ and $L_{0}$ are,
respectively, the capacitance and inductance per unit length of the
transmission line resonator; and $\phi''$ refers to the second-order
spatial derivative. Note that here we have used the continuous
variable $x$ instead of the discrete variable $j$.

In terms of the relation between the electric current and the phase
$I\left( x,t\right) =-\Phi_{0}\phi'(x,t)/(2\pi L_{0})$, the boundary
condition for this system is~\cite{Wendin}
\begin{equation}
\label{boundarycondition} I\left( 0,t\right)=0, \hspace{0.5 cm}\phi
\left(l_{0},t\right)=\phi _{s}\left( t\right),
\end{equation}
where $l_{0}$ is the length of the transmission line resonator. The
wave equation~(\ref{waveequationofTLR}) with the boundary
condition~(\ref{boundarycondition}) can be solved by assuming the
solution $\phi \left( x,t\right) =\left[ A_{1}\cos \left( kvt\right)
+A_{2}\sin \left( kvt\right) \right] \cos \left( kx\right)$. The
Euler-Lagrange equation for the phase variable $\phi _{s}$ leads to
the following dispersion equation~\cite{Wendin}
\begin{equation}
kl_{0}\tan \left( kl_{0}\right) =\left( \frac{2\pi }{\Phi
_{0}}\right) ^{2}E_{J}\left( f\right)
L_{\textrm{cav}}-\frac{2C_{s}}{C_{\textrm{cav}}}k^{2}l_{0}^{2},\label{dispersionequation}
\end{equation}
where $L_{\textrm{cav}}=L_{0}l_{0}$ and
$C_{\textrm{cav}}=C_{0}l_{0}$. The wave vectors $k$ of the resonant
modes in the transmission line resonator are the solutions of the
dispersion equation~(\ref{dispersionequation}). Since the effective
Josephson energy $E_{J}\left( f\right)$ of the SQUID is tunable
through the bias magnetic flux $\Phi_{x}$, the wave vectors $k$ can
be tuned continuously by controlling $\Phi_{x}$. This
approach~\cite{Wendin} to tune cavities could be used to tune the
frequency of either one cavity or two cavities in our proposal.

\begin{figure}[tbp]
\includegraphics[bb=108 344 440 595, width=7.8 cm]{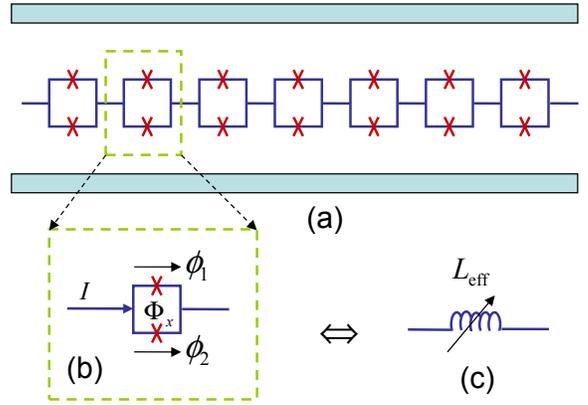}
\caption{(Color online) (a) Circuit model of a series array of
SQUIDs. (b) Circuit model of a SQUID, which is equivalent to an
effective tunable inductor in (c).} \label{fretunableTLR2.eps}
\end{figure}
\subsection{Tuning the frequency of a superconducting transmission line resonator: changing the effective inductance of a SQUID}
Following Ref.~\cite{Lehnert}, Figure~\ref{fretunableTLR2.eps}(a)
shows a device where the center of the resonator is composed of a
series array of SQUIDs. A symmetric SQUID (in
Fig.~\ref{fretunableTLR2.eps}(b)) is equivalent to an effective
tunable inductance (shown in Fig.~\ref{fretunableTLR2.eps}(c)). From
Kirchhoff's current law and the Josephson current-phase
relation~\cite{Lehnert}, then
\begin{equation}
I=I_{c}\sin\phi_{1}+I_{c}\sin\phi_{2},
\end{equation}
where $I_{c}$ is the critical current of a single Josephson
junction, $\phi_{j}$ $(j=1,2)$ are the phases across the two
Josephson junctions. Introducing new phase variables
$\phi=(\phi_{1}+\phi_{2})/2$ and $\phi_{1}-\phi_{2}=\phi_{x}$, then
$I=I_{c}\left(\phi_{x}\right) \sin\phi$, where
$I_{c}\left(\phi_{x}\right)=2I_{c}\cos(\phi_{x}/2)$. The phase can
be expressed as $\phi=\arcsin
\left[I/I_{c}\left(\phi_{x}\right)\right]$. An effective inductance
$L_{\textrm{eff}}$ can be defined~\cite{Lehnert} by the phase $\phi$
and current $I$,
\begin{eqnarray}
L_{\textrm{eff}}=\frac{\Phi _{0}\phi}{2\pi I},
\end{eqnarray}
which can be expressed as~\cite{Lehnert}
\begin{eqnarray}
L_{\textrm{eff}}\left( I,\phi_{x}\right)=\frac{\Phi _{0}}{2\pi
I}\arcsin \left( \frac{I}{I_{c}\left(\phi_{x}\right) }\right).
\end{eqnarray}
Obviously, the effective inductance of the symmetric SQUID can be
controlled through two externally controllable parameters: the
biasing current $I$ and the external biasing flux
$\Phi_{x}=\Phi_{0}\phi_{x}/(2\pi)$. Thus the resonant frequency of
the modes in a SQUID array becomes tunable because the inductance
per unit length of the center conductor of the transmission line
resonator is controllable. This approach~\cite{Lehnert} could also
be used to tune the frequency of either one cavity or two cavities
in our proposed system. This would allow the exploration of the
effect predicted here.

\subsection{Experimental implementation of our proposal}
Let us now provide some remarks on the experimental implementation
of our proposal. In our model, the key elements are the
frequency-tunable cavities, which have recently been realized
experimentally. For example, in Ref.~\cite{Delsing}, the resonant
frequency $\omega_{c}$ of a transmission line resonator was tuned
from $2\pi\times4$ GHz to $2\pi\times4.8$ GHz (i.e., $0 \leq
|\lambda\omega_{c}|\leq 2\pi\times800$ MHz). If we choose
$\omega_{c}\approx2\pi\times4$ GHz, then $0\leq\lambda\leq0.2$, if
we choose $\omega_{c}\approx2\pi\times4.8$ GHz, then
$-0.2\leq\lambda\leq0$. Similarly for $\lambda_{0}$. In principle,
the hopping coupling $J$ between two nearest-neighbor transmission
line resonators can be tuned~\cite{Liao09}. In recent experiments
(e.g., in Ref.~\cite{Houck}), the magnitude of the hopping
interaction is $J\approx2\pi\times44$ MHz $\approx0.01\omega_{c}$.
This hopping coupling $J$ can be increased by using larger
capacitors to connect two transmission line resonators. Therefore,
this study seems to be within the reach of current (or near future)
experiments.

Compared to the method using a two-level atom as a controller, in
Refs.~\cite{ZGLSN08,ZDLSN08}, the present proposal avoids photon
dissipation due to the spontaneous emission of the atom.

It should be pointed out that we have neglected the change of the
hopping coupling $J$ between the frequency-tunable cavity and its
nearest-neighbor cavities when the frequency of the FTC is tuned. In
practice, this dependence exists.

\section{\label{Sec:5}Summary}
In conclusion, we have studied controllable single-photon transport
and single-photon states in a one-dimensional coupled-cavity array
(CCA) with one or two frequency-tunable cavities (FTCs). We found
that, by adjusting the frequency of the frequency-tunable cavities,
the coherent transport of a single photon in the CCA can be
realized. We have also shown that there exist bound states in the
CCA.

For a CCA with \textit{one} FTC, when the frequency of the FTC is
larger than those of other cavities, there exists a \textit{bound}
state above the energy band of the CCA. When the frequency of the
FTC is smaller than those of other cavities, there exists a bound
state below the energy band. In these two cases, the bound states
have even parity. Once the frequency of the FTC is given, the CCA
has only one bound state. This result is different from that of a
CCA coupled with a two-level atom~\cite{ZDLSN08,ZGLSN08}, in which
there exists two bound states at the same time, one above the energy
band and the other below the energy band.

For a CCA with \textit{two} FTCs, in the two cases that the
frequencies of the two FTCs are larger or lower than those of other
cavities, there exist two \textit{bound} states, one of odd parity
and the other one of even parity. When the frequency detuning
$\lambda_{0}\omega_{c}$ of the two FTCs is very larger than the
hopping coupling $J$ between two nearest-neighbor cavities, there
exist two kinds of \textit{resonant} modes, one of odd parity, and
another one of even parity.

\begin{acknowledgments}
This work is supported in part by NSFC Grants No.~10935010 and
No.~10704023, NFRPC Grants No.~2006CB921205, No.~2007CB925204, and
NCET-08-0682. Y. X. Liu is supported by NSFC Grants No. 10975080 and
No. 60836001. F.N. acknowledges partial support from the National
Security Agency, Laboratory Physical Science, Army Research Office,
National Science Foundation Grant No. 0726909, and JSPS-RFBR
Contract No.~06-02-91200.
\end{acknowledgments}

\end{document}